\documentclass[aps,prb,showpacs,twocolumn,amsmath,amssymb]{revtex4-2}

\usepackage{amsmath}    
\usepackage{amssymb}
\usepackage{graphicx}   
\usepackage{graphics}
\usepackage{dcolumn}
\usepackage{bm}

\usepackage{multirow}

\usepackage{latexsym}
\usepackage{amssymb}
\usepackage{amsmath}

\usepackage{color}
\usepackage{epstopdf}

\usepackage[normalem]{ulem}

\begin{document}
\title{Lattice elasticity of blue phases in cholesteric liquid crystals}
\author{V.~A.~Chizhikov$^{\phantom{;}a,b}$}
\email{chizhikov@crys.ras.ru}
\author{A.~V.~Mamonova$^{\phantom{;}a,c}$}
\author{V.~E.~Dmitrienko$^{\phantom{;}a,c}$}
\email{dmitrien@crys.ras.ru}
\address{$^a$Shubnikov Institute of Crystallography of the Kurchatov Complex of Crystallography and Photonics of the National Research Center ``Kurchatov Institute'', Moscow, 119333 Russia,}
\address{$^b$MIREA --- Russian Technological University (Institute of Radio Electronics and Informatics), Moscow, 119454 Russia,}
\address{$^c$Osipyan Institute of Solid State Physics of the Russian Academy of Sciences, Chernogolovka, Moscow Region, 142432 Russia}

\begin{abstract} 
New theoretical approaches have been developed for studying and quantitatively describing the elastic properties of cubic blue phases in cholesteric liquid crystals. Within the framework of the Landau–de Gennes theory, using the simplest blue phase with spatial group $O^5$ ($I432$) as an example, calculations of the bulk modulus and two shear moduli were performed depending on the chirality strength and temperature below the crystallization point from isotropic liquid. It is shown that the used approximations of rigid tensors and free helicoids give qualitatively similar results but differ noticeably quantitatively, therefore further experimental studies and numerical modeling of blue phase elasticity are necessary.
\end{abstract}
\maketitle


\section{Introduction}
\label{sec:intro}

Liquid crystals (LC) are convenient objects for studying the influence of various types of ordering on elastic properties. These can be purely orientational ordering as in nematic crystals, as well as various combinations of orientational and one-, two-, and three-dimensional translational ordering. Blue phases are unique in this sequence because, while remaining liquid at the molecular level (the centers of gravity of molecules are not fixed in space), they form crystals exclusively through the creation of a three-dimensionally periodic structure in molecular orientations, and such crystals possess shear elastic moduli \cite{Belyakov1985,Wright1989,Collings2014}, albeit of very small magnitude. Experimentally, the elastic moduli of blue phases have been studied in several works since 1984 \cite{Clark1984,Cladis1984,Kleiman1984,Gleeson2015}, however, these works used polycrystalline samples, which do not allow obtaining complete quantitative information.

The development of possible practical applications of blue phases has been ongoing for a long time and in various directions \cite{Rahman2015,Khosla2020,Bagchi2023}. In particular, it has been demonstrated that displays based on blue phases have significantly higher performance than those currently in use, but their application is still limited by the need for high control voltage, which in turn depends on the elasticity of blue phases.

It should be noted that the problem of elasticity of structures with macroscopic periods extends far beyond blue phases. Similar problems arise for three-dimensional, two-dimensional, and onedimensional structures forming in liquid crystals due to very different physical reasons \cite{Dolganov2022,Baklanova2023,Blanc2023}. Moreover, there appears to be a close analogy between the behavior of cholesterics during the transition from isotropic liquid to blue phases \cite{Brazovskii1975,Brazovskii1978} and phase transitions in spiral magnets \cite{Stishov2023}. The similarity of these physical phenomena is discussed in detail in a recent review \cite{Demishev2024}.

All this compels us to return to studying the physical properties of blue phases of liquid crystals at a new level of understanding of the underlying problems and possibilities for their solution, in order to later use the obtained results for other structures with macroscopic periods, such as skyrmion magnetic lattices, periodically ordered confocal domains, etc.

In this work, we consider elastic properties using the simplest phase with the cubic space group $O^5$ ($I432$) as an example, which is formed by six equivalent helicoids directed along the two-fold axes  $\left<110\right>$. On one hand, this phase is reliably predicted within the framework of the Landau–de Gennes theory for liquid crystals with high chirality \cite{Belyakov1985,Wright1989}, but it has not yet been observed experimentally either because the chirality of real crystals is not sufficiently high, or because in highly chiral crystals, the periods and consequently the Bragg reflections lie in the far ultraviolet region, which is very difficult to observe. On the other hand, it is also interesting because it allows further advancement in analytical calculations of physical properties without resorting to cumbersome computer modeling, which is practically inevitable for the actually observed phases $O^2$ and $O^8$, which we plan to consider in the future.

\section{Order parameter and free energy of chiral LC$\mathbf{s}$}
\label{sec:order}

Usually, as an order parameter for nematic and cholesteric LCs, which can be considered locally uniaxial, a scalar orientational ordering parameter is used, characterizing the averaged angle of local molecular disorientation: $S = (3\left< \cos^2\vartheta \right>-1)/2$, and the local direction of average molecular orientation is described by a unit vector $\mathbf{n}$, called the {\it director}. Unlike uniaxial nematics and cholesterics, blue phases are essentially locally biaxial at least in part of the volume, and as an order parameter for the phenomenological description of the phase transition from isotropic liquid to crystalline blue phases, a traceless symmetric tensor $\hat{Q}$ is used, having five independent components that describe both local biaxial ordering and its orientation in space \cite{Belyakov1985,Wright1989}. For example, the order parameter can be the anisotropic part of the local dielectric permittivity tensor. Following the notation of the classical review article by Wright and Mermin \cite{Wright1989}, which is based on the works of Grebel, Hornreich, and Strickman \cite{Grebel1983}, the Landau--de Gennes free energy density can be divided into gradient and bulk parts:
\begin{equation}
	\label{eq:free-energy}
	\begin{array}l
		f = f_\mathrm{grad} + f_\mathrm{bulk} , \vspace{0.2cm} \\
		f_\mathrm{grad} = \tfrac14 K_1 [(\boldsymbol{\nabla} \times \hat{Q})_{\alpha\beta} + 2 q_0 Q_{\alpha\beta}]^2 + \tfrac14 K_0 [(\boldsymbol{\nabla} \cdot \hat{Q})_\alpha]^2 , \vspace{0.2cm} \\
		f_\mathrm{bulk} = c \phantom{.} \mathrm{Tr}(\hat{Q}^2) - \sqrt{6} b \phantom{.} \mathrm{Tr}(\hat{Q}^3) + a \phantom{.} [\mathrm{Tr}(\hat{Q}^2)]^2 .
	\end{array}
\end{equation}
Here
\[
\begin{array}c
(\boldsymbol{\nabla} \times \hat{Q})_{\alpha\beta} = \epsilon_{\alpha\mu\nu} \nabla_\mu Q_{\nu\beta} , \vspace{0.2cm} \\
(\boldsymbol{\nabla} \cdot \hat{Q})_\alpha = \nabla_\mu Q_{\mu\alpha} ,
\end{array}
\]
$K_0$ and $K_1$ are elastic moduli of the liquid crystal, wave number $q_0$ characterizes the chirality of LC (cholesteric pitch; for nematic $q_0 = 0$), and coefficient $c$ at the quadratic term of the bulk part depends on temperature and changes sign near the transition point from isotropic phase to cholesteric. At $q_0 \neq 0$ it is convenient to make the free energy expression dimensionless by introducing notations
\[
\begin{array}c
	\varphi = (a^3 / b^4) f , \phantom{xx} \tau = (a / b^2) c, \vspace{0.2cm} \\
	\hat{\chi} = (a / b) \hat{Q} , \phantom{xx} r_\mathrm{new} = 2 |q_0| r_\mathrm{old},
\end{array}
\]
and the free energy density transforms to the form
\begin{equation}
	\label{eq:varphi}
	\begin{array}l
		\varphi = \varphi_\mathrm{grad} + \varphi_\mathrm{bulk} , \vspace{0.2cm} \\
		\varphi_\mathrm{grad} = \kappa^2 \{[(\boldsymbol{\nabla} \times \hat{\chi})_{\alpha\beta} \pm \chi_{\alpha\beta}]^2 + \eta [(\boldsymbol{\nabla} \cdot \hat{\chi})_\alpha]^2 \} , \vspace{0.2cm} \\
		\varphi_\mathrm{bulk} = \tau \mathrm{Tr}(\hat{\chi}^2) -\sqrt{6} \mathrm{Tr}(\hat{\chi}^3) + [\mathrm{Tr}(\hat{\chi}^2)]^2 .
	\end{array}
\end{equation}
Here $\kappa = |q_0 / b| \sqrt{a K_1}$ is a dimensionless parameter determining the weight ratio between gradient and bulk parts of free energy, and $\eta = K_0 / K_1$. Parameter $\kappa$ is often called {\it chirality} \cite{Grebel1983}, since only gradient energy is not invariant with respect to spatial coordinate inversion. The sign of chirality is determined by choosing between plus and minus in the first square brackets in $\varphi_\mathrm{grad}$. Henceforth, we will always choose the plus sign, corresponding to a righthanded helix.

Note that the gradient part of free energy in (\ref{eq:free-energy}) contains only two bulk elastic moduli: $K_0$ and $K_1$, rather than three as in Frank's energy, which describes the elastic properties of nematics and cholesterics using the director $\mathbf{n}$:
\begin{equation}
	\label{eq:Frank-energy}
	f_\mathrm{F} = \tfrac12 K_1^\mathrm{F} (\mathrm{div} \mathbf{n})^2 + \tfrac12 K_2^\mathrm{F} (\mathbf{n} \cdot \mathrm{curl} \mathbf{n} + q)^2 + \tfrac12 K_3^\mathrm{F} [\mathbf{n} \times \mathrm{curl} \mathbf{n}]^2 .
\end{equation}
At temperatures below the transition point from isotropic liquid to ordered nematic or cholesteric phase, a transition is possible from tensor formalism to description in terms of director $\mathbf{n}$ by replacing the traceless symmetric tensor $||Q_{\alpha\beta}||$ with a uniaxial traceless tensor $S ||n_{\alpha}n_{\beta} - \tfrac13\delta_{\alpha\beta}||$. Such description works well throughout the LC volume, except for small regions around defects. Assuming $S = const$, it's easy to obtain from $f_\mathrm{grad}$ Frank's energy (\ref{eq:Frank-energy}) with the following relation between elastic moduli:
\[
K_1^\mathrm{F} = K_3^\mathrm{F} = \tfrac12 S^2 (K_0 + K_1) , \phantom{x} K_2^\mathrm{F} = S^2 K_1 .
\]
Thus, in the theory based on expression (\ref{eq:free-energy}) for free energy, two Frank moduli coincide. The experimentally observed difference between them may result from, for example, gradient terms of higher orders in tensor $\hat{Q}$, but these effects are not considered here. Note that when $K_0 = K_1$ all Frank bulk moduli are equal to each other; this approximation is called {\it one-parameter}.

The competition between gradient and bulk contributions to free energy leads to a non-trivial phase diagram of cholesteric LCs. Thus, the minimum of gradient energy corresponds to biaxial helicoids formed by harmonics of the form
\begin{equation}
	\label{eq:helix}
	\hat{\chi} \sim (\mathbf{m}_1 - i \mathbf{m}_2) \otimes (\mathbf{m}_1 - i \mathbf{m}_2) \exp(i \mathbf{k} \cdot \mathbf{r}) ,
\end{equation}
where the wave vector $\mathbf{k}$ has unit length ($k = 1$), and vectors $\mathbf{m}_1$ and $\mathbf{m}_2$ form an orthonormal basis in the perpendicular to $\mathbf{k}$ plane ($[\mathbf{m}_1 \times \mathbf{m}_2] = \mathbf{k}$ for right-handed helicoid). On the other hand, the bulk energy is minimized by a uniaxial tensor $\hat{\chi}$, and at temperatures significantly below the transition point from isotropic liquid to cholesteric phase, the uniaxial helicoidal structure, conveniently described by the director, becomes energetically favorable. Near the transition point, as a result of the competition between gradient and bulk energies, spatial lattices are formed by superposition of several biaxial helicoids of form (\ref{eq:helix}) with crossed wave vectors. Such lattice phases can exist in a very narrow temperature range, and due to their property of scattering light of specific wavelengths, they are called {\it blue phases}.

Several different blue phases have been experimentally observed. In a typical cholesteric, during the transition from isotropic liquid to uniaxial helicoidal one, three phases successively emerge in a narrow temperature range, replacing each other: BP~III (also called {\it blue fog}), BP~II, and BP~I. The blue fog is amorphous, while the other two possess cubic symmetry. In electric fields, a hexagonal phase was observed, largely similar to the skyrmion A phase in cubic helimagnets \cite{Chizhikov2021}. Additionally, the possibility of other phases with different symmetries, such as icosahedral, has been discussed. As mentioned above, all blue phases are formed by the superposition of crossed (right) helicoids of type (\ref{eq:helix}) with wave vectors $\mathbf{k}$ of equal or at least similar length to minimize the gradient energy as much as possible. The method of crossing (angles between wave vectors, phases of helicoids) should ideally minimize the bulk energy. Competition between gradient and bulk energies may lead to some distortion of the structure (tensors $\hat{\chi}_\mathbf{k}$ of general form, wave vectors with $k \neq 1$), but the contribution to energy from such distortions remains insignificant.

To describe blue phases formed by crossed helicoids, especially in the case of periodic structures, it is convenient to switch to the Fourier representation for the order parameter:
\begin{equation}
	\label{eq:Fourier}
	\hat{\chi}(\mathbf{r}) = \sum_\mathbf{k} \hat{\chi}_\mathbf{k} \exp(i \mathbf{k} \cdot \mathbf{r}) ,
\end{equation}
where, due to the reality of the field $\hat{\chi}(\mathbf{r})$, $\hat{\chi}_{-\mathbf{k}} = \hat{\chi}^\ast_\mathbf{k}$. Then the volume-averaged densities of gradient and bulk energies (\ref{eq:varphi}) will take the form
\begin{equation}
	\label{eq:phigrad-fourier}
	\begin{array}{l}
		\left< \varphi_\mathrm{grad} \right> = \kappa^2 \sum_\mathbf{k} \{ (k^2 + 1) \mathrm{Tr}(\hat{\chi}_\mathbf{k} \cdot \hat{\chi}^\ast_\mathbf{k}) + \vspace{0.2cm} \\ 
		\phantom{xx} + (\eta - 1) \mathbf{k} \cdot \hat{\chi}_\mathbf{k} \cdot \hat{\chi}^\ast_\mathbf{k} \cdot \mathbf{k} + 2 i \epsilon_{\alpha\beta\gamma} k_\alpha (\hat{\chi}_\mathbf{k} \cdot \hat{\chi}^\ast_\mathbf{k})_{\beta\gamma} \} ,
	\end{array}
\end{equation}
\begin{equation}
	\label{eq:phibulk-fourier}
	\begin{array}{l}
		\left< \varphi_\mathrm{bulk} \right> = \tau \sum_\mathbf{k} \mathrm{Tr}(\hat{\chi}_\mathbf{k} \cdot \hat{\chi}^\ast_\mathbf{k}) - \vspace{0.2cm} \\
		\phantom{xx}  - \sqrt{6} \sum_{\mathbf{k}_1 + \mathbf{k}_2 + \mathbf{k}_3 = 0} \mathrm{Tr}(\hat{\chi}_{\mathbf{k}_1} \cdot \hat{\chi}_{\mathbf{k}_2} \cdot \hat{\chi}_{\mathbf{k}_3}) + \vspace{0.2cm} \\
		\phantom{xx} + 2 \sum_{\mathbf{k}_1 + \mathbf{k}_2 + \mathbf{k}_3 + \mathbf{k}_4 = 0} \mathrm{Tr}(\hat{\chi}_{\mathbf{k}_1} \cdot \hat{\chi}_{\mathbf{k}_2} \cdot \hat{\chi}_{\mathbf{k}_3} \cdot \hat{\chi}_{\mathbf{k}_4}) ,
	\end{array}
\end{equation}
respectively. It should be noted that such consideration of the phase transition from isotropic liquid to crystalline blue phases is in many ways analogous to the theory of weak crystallization \cite{Kats1993}, which is sometimes applicable to ordinary crystals, for example, when this transition is a weak first-order transition.

In the next section, we will show how to minimize the free energy of the blue phase described by expressions (\ref{eq:phigrad-fourier}), (\ref{eq:phibulk-fourier}), using the example of a hypothetical cubic phase $O^5$.

\section{Blue phase $O^5$}
\label{sec:O5}

Let us consider a hypothetical blue phase with cubic space group $O^5$ ($I432$), formed by twelve harmonics of type (\ref{eq:helix}) with wave vectors directed along crystallographic directions $\left< 110 \right>$. Choosing an orthonormal basis in the plane perpendicular to vector $\mathbf{k} = (kk0) / \sqrt{2}$, $\mathbf{m}_1 = (001)$, $\mathbf{m}_2 = (1\bar{1}0) / \sqrt{2}$, we find from (\ref{eq:helix}):
\begin{equation}
	\label{eq:chi110e}
	\hat{\chi}_{(kk0) / \sqrt{2}} = \left(
	\begin{array}{ccc}
		- e & e & - i \sqrt{2} e \\
		e & - e & i \sqrt{2} e \\
		- i \sqrt{2} e & i \sqrt{2} e & 2 e
	\end{array}
	\right) .
\end{equation}
The remaining tensors $\hat{\chi}_\mathbf{k}$ are obtained from $\hat{\chi}_{(kk0) / \sqrt{2}}$ by symmetry transformations of the cubic class $O$. Note that since tensors $\hat{\chi}_\mathbf{k}$ and $\hat{\chi}_{-\mathbf{k}} = \hat{\chi}^\ast_\mathbf{k}$ are connected by a 4-fold rotation axis, coefficient $e$ is a real number, meaning that the phases of helicoids are determined by the space group $O^5$.

The average density of gradient energy (\ref{eq:phigrad-fourier})
\begin{equation}
	\label{eq:varphigradO5}
	\left< \varphi_\mathrm{grad} \right> = 192 \kappa^2 e^2 (k - 1)^2
\end{equation}
becomes zero at $k = 1$, and expression (\ref{eq:phibulk-fourier}) gives us the value of total energy density
\begin{equation}
	\label{eq:varphiO5}
	\left< \varphi \right> = \left< \varphi_\mathrm{bulk} \right> =  192 \tau e^2 - 1104 \sqrt{6} e^3 + 47904 e^4
\end{equation}
as a function of ``temperature'' $\tau$ and the single order parameter $e$.

Below the first-order phase transition point
\begin{equation}
	\label{eq:tauc}
	\tau_c = \tfrac{1587}{7984} \approx 0.19877 ,
\end{equation}
energy (\ref{eq:varphiO5}) is minimized by the order parameter value
\begin{equation}
	\label{eq:e}
	e = \frac{69 \sqrt{6} + \sqrt{28566 - 127744 \tau}}{7984} ,
\end{equation}
which at the transition point experiences a jump from zero to
\begin{equation}
	\label{eq:ec}
	e_c = \tfrac{23 \sqrt{6}}{1996} \approx 0.028226 .
\end{equation}

It was shown that the blue phase $O^5$ can exist in cholesteric LCs with sufficiently high chirality ($\kappa > \frac32$), in which gradient energy prevails over bulk energy, making possible the emergence of doubletwist structures and, consequently, with increased biaxiality. Calculations for this case show that phase $O^5$ has lower energy than other candidate phases near $\tau = 0$. However, in cholesterics with low chirality ($\kappa < \frac32$), which are typically observed in experiments, phase $O^5$ loses to cubic phases with space groups $O^2$ ($P4_232$) and $O^8$ ($I4_132$). It has been experimentally established that these are the symmetry groups possessed by phases BP~I ($O^8$) and BP~II ($O^2$). It also turns out that at sufficiently low chirality ($\kappa < 0.46945$) phase $O^5$ loses to the helicoidal phase at any temperature below $\tau_c$.

\section{Elastic energy of the deformed blue phase}
\label{sec:elasticity}

When mechanical stress is applied to a crystal, elastic deformations arise, described by a symmetric tensor $\hat{\varepsilon}$. The elastic strain energy density is expressed through the tensor $\hat{\varepsilon}$ as follows:
\begin{equation}
	\label{eq:varphied}
	\varphi_\mathrm{e.d.} = \tfrac12 \lambda_{\alpha\beta\gamma\delta} \varepsilon_{\alpha\beta} \varepsilon_{\gamma\delta} ,
\end{equation}
where $\hat{\lambda}$ is the elasticity tensor, which in the case of a cubic crystal has three independent components: $\lambda_{xxxx}$, $\lambda_{xxyy}$ and $\lambda_{xyxy}$. To calculate them, it is necessary to determine what real structural distortions occur in the blue phase during deformation described by the tensor $\hat{\varepsilon}$. Thus, it is obvious that the helicoid wave vectors $\mathbf{k}$, which make up the crystal's reciprocal lattice, transform in a strictly determined way. At the same time, the tensors $\hat{\chi}_\mathbf{k}$ relax to a new state that minimizes energy (\ref{eq:varphi}) for the distorted reciprocal lattice. This means that the elastic energy density can be expressed through changes in wave vectors $\mathbf{k}$ and tensors $\hat{\chi}_\mathbf{k}$:
\begin{equation}
	\label{eq:varphied2}
	\begin{array}{l}
		\varphi_\mathrm{e.d.} = \displaystyle \tfrac12 \sum_{i,j} \frac{\partial^2 \left< \varphi \right>}{\partial k_{i,\alpha} \partial k_{j,\beta}} \Delta k_{i,\alpha} \Delta k_{j,\beta} - \vspace{0.2cm} \\
		\phantom{xxxx} \displaystyle - \tfrac12 \sum_{i,j} \frac{\partial^2 \left< \varphi \right>}{\partial \chi_{i,\alpha\beta} \partial \chi_{j,\gamma\delta}} \Delta \chi_{i,\alpha\beta} \Delta \chi_{j,\gamma\delta} .
	\end{array}
\end{equation}
Comparing equations (\ref{eq:varphied}) and (\ref{eq:varphied2}), it is evident that it is sufficient to consider only first-order contributions with respect to the strain tensor in $\Delta \mathbf{k}$ and $\Delta \hat{\chi}_\mathbf{k}$.

First, let us show how the wave vectors of helicoids that make up the blue phase of cholesteric LC deform. Consider a uniform deformation where a crystal point with coordinate $\mathbf{r}$ shifts by distance $\mathbf{u} = \hat{u} \cdot \mathbf{r}$, moving to a new position
\begin{equation}
	\label{eq:rprime}
	\mathbf{r}^\prime = (1 + \hat{u}) \cdot \mathbf{r} .
\end{equation}
Tensor $\hat{u}$ is related to the strain tensor in a nonlinear way:
\begin{equation}
	\label{eq:varepsilon-u}
	\hat{\varepsilon} = \tfrac12 (\hat{u} + \hat{u}^T + \hat{u}^T \cdot \hat{u}) .
\end{equation}
Note, however, that the antisymmetric part of tensor $\hat{u}$ corresponds to spatial rotation of the crystal, not its deformation. Excluding this rotation, we can consider $\hat{u}^T = \hat{u}$, without losing generality. Then
\begin{equation}
	\label{eq:varepsilon-u2}
	\hat{\varepsilon} = \hat{u} + \frac12 \hat{u}^2 \approx \hat{u} 
\end{equation}
and
\begin{equation}
	\label{eq:rprime2}
	\mathbf{r}^\prime = (1 + \hat{\varepsilon}) \cdot \mathbf{r} .
\end{equation}
For the helicoid phase $\mathbf{k} \cdot \mathbf{r}$ to remain unchanged at each point, the new wave vector in the first approximation will be equal to
\begin{equation}
	\label{eq:kprime}
	\mathbf{k}^\prime =  (1 - \hat{\varepsilon}) \cdot \mathbf{k} ,
\end{equation}
from which
\begin{equation}
	\label{eq:Deltak}
	\Delta \mathbf{k} = \mathbf{k}^\prime - \mathbf{k} = - \hat{\varepsilon} \cdot \mathbf{k} .
\end{equation}

The change in tensors $\hat{\chi}_\mathbf{k}$ can be found by minimizing the energy density (\ref{eq:varphi}) at given wave vectors $\mathbf{k}^\prime$. However, some approximate results can be obtained within simple models considered in the following subsections.

\subsection{Rigid Tensor $\hat{\chi}$ Approximation}
\label{sec:hard-tensor}

\begin{figure}[h]
	\begin{center}
		\includegraphics[width=8cm]{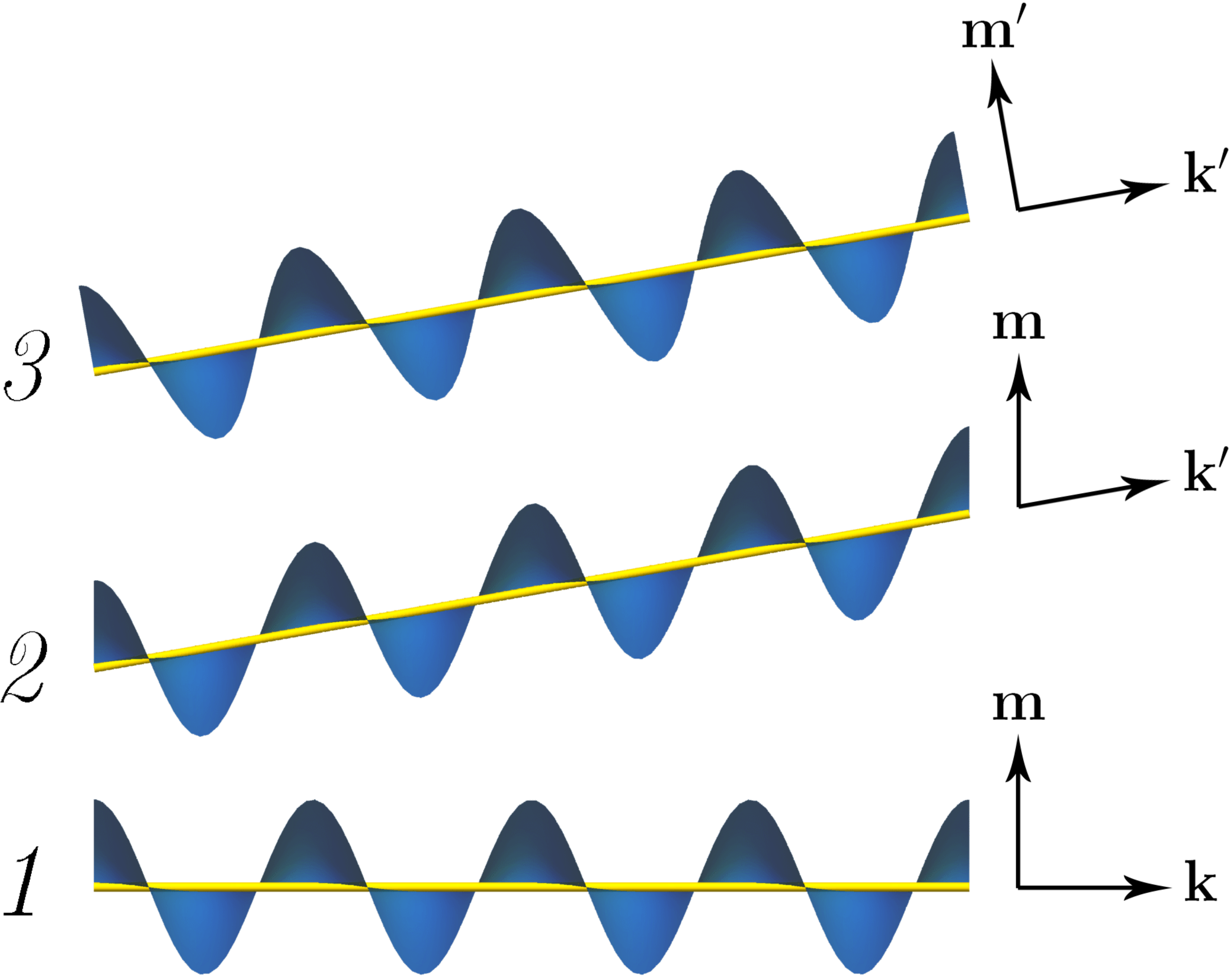}
		\caption{\label{fig:helices} Wave vectors and rotation planes of helicoids: {\it 1} --- helicoid in undeformed crystal, {\it 2} --- rigid tensor approximation $\hat{\chi}$, {\it 3} --- free helicoid approximation
		}
	\end{center}
\end{figure}

Let us assume that during deformation, a large bulk contribution to the free energy rigidly fixes the cholesteric LC order parameter in its initial state, that is
\begin{equation}
	\label{eq:chichi}
	\hat{\chi}^\prime(\mathbf{r}^\prime) = \hat{\chi}(\mathbf{r}) .
\end{equation}
This means that the Fourier components of the tensor $\hat{\chi}(\mathbf{r})$, given by expression (\ref{eq:chi110e}), do not change. Accordingly, the bulk energy density is preserved: $\Delta \left< \varphi_\mathrm{bulk} \right> = 0$, and the elastic deformation energy is completely determined by the change in wave vectors $\mathbf{k}$ in the gradient energy (\ref{eq:phigrad-fourier}): $\varphi_\mathrm{e.d.} = \Delta \left< \varphi_\mathrm{grad} \right>$. Note that the wave vector $\mathbf{k}^\prime$ of the deformed structure is no longer perpendicular to the rotation plane of the tensor $\hat{\chi}_\mathbf{k}$, defined by vectors $\mathbf{m}_1$ and $\mathbf{m}_2$ (see Fig.~\ref{fig:helices}). This means that the helical spirals forming the blue phase acquire a small cycloidal component. Substituting wave vectors $\mathbf{k}^\prime$ for general deformation into (\ref{eq:phigrad-fourier}) allows finding the components of the elasticity tensor in the rigid tensor approximation:
\begin{equation}
	\label{eq:lambda1}
	\left( \begin{array}{c} \lambda_{1,xxxx} \\ \lambda_{1,xxyy} \\ \lambda_{1,xyxy} \end{array} \right) = 16 \kappa^2 e^2 \left( \begin{array}{c} 6 + 2 \eta \\ 1 - \eta \\ 3 + \eta \end{array} \right) .
\end{equation}
It is interesting to note that the component $\lambda_{xxyy}$ takes negative values at $\eta > 1$, which, however, should not lead to negativity of elastic deformation energy (the latter would mean that the crystal is unstable with respect to spontaneous deformation). Indeed, the positive definiteness of the quadratic form (\ref{eq:varphied}) for non-shear deformations is determined by conditions
\begin{equation}
	\label{eq:conditions}
	\begin{array}{l}
		\lambda_{xxxx} > |\lambda_{xxyy}| , \vspace{0.2cm} \\
		\lambda_{xxxx} + 2 \lambda_{xxyy} > 0 ,
	\end{array}
\end{equation}
which are satisfied at positive values of $\eta$. The last condition of (\ref{eq:conditions}) is nothing but the requirement for positivity of the bulk modulus of the blue phase
\begin{equation}
	\label{eq:K}
	K = \tfrac13 (\lambda_{xxxx} + 2 \lambda_{xxyy}) .
\end{equation}

\subsection{Free Helicoid Approximation}
\label{sec:free-helicoid}

In contrast to the rigid tensor approximation, an alternative model can be proposed in which the field $\hat{\chi}(\mathbf{r})$ is defined by a superposition of ideal helicoids that do not interact with each other. This approach, proposed in \cite{Dmitrienko1986}, is strictly speaking applicable to cholesteric LCs with high chirality ($\kappa \gg 1$), in which changes in the bulk part of free energy can be neglected. In this case, the elastic deformation energy is determined by the change in periods of the helicoids comprising the blue phase.

The wave number of each helicoid is inversely proportional to its period and changes as follows under small deformation
\begin{equation}
	\label{eq:DeltaAbsk}
	k^\prime - k = - \frac{\mathbf{k} \cdot \hat{\varepsilon} \cdot \mathbf{k}}k .
\end{equation}
In the undistorted phase $O^5$, $k = 1$ corresponds to zero gradient energy, and after deformation, the gradient energy density of one harmonic equals
\begin{equation}
	\label{eq:phi1h}
	\varphi_\mathrm{1h} = 16 \kappa^2 e^2 (k^\prime - 1)^2
\end{equation}
(cf. with expression (\ref{eq:varphigradO5}) for phase $O^5$, composed of 12 harmonics). Summing (\ref{eq:phi1h}) over all harmonics, we calculate the elastic deformation energy density
\begin{equation}
	\label{eq:varphied3}
	\varphi_\mathrm{e.d.} = 16 \kappa^2 e^2 \sum_{\mathbf{k}} \frac{(\mathbf{k} \cdot \hat{\varepsilon} \cdot \mathbf{k})^2}{k^2} .
\end{equation}
Comparing (\ref{eq:varphied3}) and (\ref{eq:varphied}), we find the elasticity tensor
\begin{equation}
	\label{eq:lambda2}
	\lambda_{\alpha\beta\gamma\delta} = 32 \kappa^2 e^2 \sum_{\mathbf{k}} \frac{k_\alpha k_\beta k_\gamma k_\delta}{k^2} ,
\end{equation}
symmetric with respect to permutations of all indices.

For phase $O^5$, after summing over all vectors of type $(110) / \sqrt{2}$, we obtain
\begin{equation}
	\label{eq:lambda3}
	\left( \begin{array}{c} \lambda_{2,xxxx} \\ \lambda_{2,xxyy} \\ \lambda_{2,xyxy} \end{array} \right) = 16 \kappa^2 e^2 \left( \begin{array}{c} 4 \\ 2 \\ 2 \end{array} \right) .
\end{equation}
We can compare this with result (\ref{eq:lambda1}) obtained in the rigid tensor approximation. First, we note that both approximations give the same value for the bulk modulus of elasticity:
\[
K = \frac{128}3 \kappa^2 e^2 ,
\]
since uniform expansion/compression always affects only the pitch of helicoids, leaving the Fourier components $\hat{\chi}_\mathbf{k}$ unchanged. For deformation of any other type, the free helicoid model always gives lower elastic deformation energy than the rigid tensor approximation. In principle, this should indicate that deformation by the free helicoid type is energetically more favorable. However, such comparison is not entirely legitimate since it does not take into account changes in the bulk energy of the blue phase.

\subsection{Accounting for bulk energy in the free helicoid approximation}
\label{sec:bulk-energy}

In order to estimate how large the chirality $\kappa$ should be for the applicability of the free helicoid approximation, it is necessary to consider the contribution of bulk energy to the elastic deformation energy. Note that in this model, the phases of helicoids, which do not affect the gradient energy but are important for calculating the bulk contribution to energy, have not been taken into account yet. Let's fix the phase by assuming, for example, that the rotation vector is perpendicular to the helicoid axis. This means that the rotation matrix for a helicoid with wave vector $\mathbf{k}$ has the form:
\begin{equation}
	\label{eq:R}
	R(\mathbf{k},\hat{\varepsilon}) = 1 + \frac{\mathbf{k} \otimes (\hat{\varepsilon} \cdot \mathbf{k}) -  (\hat{\varepsilon} \cdot \mathbf{k}) \otimes \mathbf{k}}{k^2} .
\end{equation}
The Fourier component of the order parameter associated with this helicoid changes as
\begin{equation}
	\label{eq:chirotation}
	\hat{\chi}^\prime_{\mathbf{k}^\prime} = R(\mathbf{k},\hat{\varepsilon}) \cdot \hat{\chi}_{\mathbf{k}} \cdot R^{-1}(\mathbf{k},\hat{\varepsilon}) .
\end{equation}
Substituting (\ref{eq:chirotation}) into expression (\ref{eq:phibulk-fourier}) for the bulk energy density, we find the contribution to the elasticity tensor from interacting helicoids:
\begin{equation}
	\label{eq:lambda4-bulk}
	\left( \begin{array}{c} \Delta \lambda_{xxxx} \\ \Delta \lambda_{xxyy} \\ \Delta \lambda_{xyxy} \end{array} \right) = \left( \begin{array}{c} 48 \sqrt{6} e^3 + 26496 e^4 \\ -24 \sqrt{6} e^3 - 13248 e^4 \\ 528 \sqrt{6} e^3 - 2176 e^4 \end{array} \right) .
\end{equation}
First of all, let's note the equality
\[
\Delta \lambda_{xxxx} + 2 \Delta \lambda_{xxyy} = 0
\]
confirming the above observation that the bulk energy does not change under uniform expansion/compression. Moreover, the negativity of $\Delta \lambda_{xxyy}$ does not lead to instability, since inequalities (\ref{eq:conditions}) are still satisfied. In contrast, the negativity of $\Delta \lambda_{xyxy}$ at $e > 0.59436$ indicates the instability of phase $O^5$ with respect to shear deformations at low temperatures.

Comparing (\ref{eq:lambda3}) with (\ref{eq:lambda4-bulk}) at the critical point  $\tau_c$ ($e_c = 0.028226$), we determine the applicability condition for the free helicoid approximation. For example, the inequality $\lambda_{xyxy} \gg \Delta \lambda_{xyxy}$ leads to the following condition for chirality: $\kappa^2 \gg 1.0866$. Thus, as assumed, the free helicoid model works well for high chirality. It can also be assumed that at low chirality ($\kappa < \tfrac32$) the changes in tensors  $\hat{\chi}_\mathbf{k}$ will represent something intermediate between what the rigid tensor approximation and the free helicoid approximation give, and such an interpolation model is considered in the next section.

\subsection{Interpolation model}
\label{sec:interpolation}

For a quantitative description of the intermediate case between the rigid tensor approximation and the free helicoid approximation, expression (\ref{eq:R}) can be modified as follows:
\begin{equation}
	\label{eq:R2}
	R(\mathbf{k},\hat{\varepsilon}) = 1 + \alpha(\mathbf{k},\hat{\varepsilon}) \frac{\mathbf{k} \otimes (\hat{\varepsilon} \cdot \mathbf{k}) -  (\hat{\varepsilon} \cdot \mathbf{k}) \otimes \mathbf{k}}{k^2} ,
\end{equation}
by introducing a coefficient $0 \leq \alpha(\mathbf{k},\hat{\varepsilon}) \leq 1$, that changes the rotation angle of the helicoid rotation plane. Then, if for all helicoids $\alpha$ are equal to zero, the model coincides with the rigid tensor approximation, and when all $\alpha$ are equal to one, it coincides with the free helicoid approximation. The coefficients $\alpha(\mathbf{k},\hat{\varepsilon})$ are not arbitrary but are determined from the condition of minimum elastic deformation energy and thus depend on both chirality and temperature.

In the case of general deformation, the number of variables $\alpha$ to be determined equals the number of helicoids: for phase $O^5$ it equals six. However, the task can be significantly simplified by using specific types of deformations that preserve certain elements of the point symmetry of the blue phase. Let's take, for example, the deformation tensor
\begin{equation}
	\label{eq:deformation001}
	\hat{\varepsilon} = \left(
	\begin{array}{ccc}
		0 & 0 & 0 \\
		0 & 0 & 0 \\
		0 & 0 & \varepsilon_{zz}
	\end{array}
	\right) ,
\end{equation}
corresponding to extension/compression along the $z$ crystal axis. The elastic deformation energy density in this case equals
\begin{equation}
	\label{eq:phi001}
	\varphi_\mathrm{e.d.} = \tfrac12 \lambda_{xxxx} \varepsilon_{zz}^2 ,
\end{equation}
which allows direct calculation of the component $\lambda_{xxxx}$ of the elasticity tensor. The point symmetry of the phase decreases from $432$ to $422$, and the set of six initially equivalent helicoids splits into two groups. Two helicoids of the first group, with axes along crystallographic directions $[110]$ and $[1\bar{1}0]$, lie perpendicular to the $z$ axis, and the rotations of their rotation planes are automatically zero. The other four helicoids: $[011]$, $[0\bar{1}1]$, $[101]$, $[\bar{1}01]$ --- are connected by a 4-fold rotation axis, and due to symmetry, the coefficient $\alpha$ is the same for them. Thus, the task of minimizing energy density (\ref{eq:phi001}) reduces to solving a quadratic equation for the number $\alpha$.

Another component of the elastic tensor, $\lambda_{xxyy}$, can be calculated using the expression for the bulk modulus of elasticity (\ref{eq:K}):
\begin{equation}
	\label{eq:lambdaxxyy}
	\lambda_{xxyy} = \tfrac32 K - \tfrac12 \lambda_{xxxx} = 64 \kappa^2 e^2 - \tfrac12 \lambda_{xxxx} .
\end{equation} 

To calculate the last component of the elastic tensor $\lambda_{xyxy}$ we use the deformation
\begin{equation}
	\label{eq:deformation111}
	\hat{\varepsilon} = \left(
	\begin{array}{ccc}
		0 & \varepsilon_{xy} & \varepsilon_{xy} \\
		\varepsilon_{xy} & 0 & \varepsilon_{xy} \\
		\varepsilon_{xy} & \varepsilon_{xy} & 0
	\end{array}
	\right) ,
\end{equation}
corresponding to tension/compression along the $[111]$ axis of the crystal. The elastic strain energy density in this case equals
\begin{equation}
	\label{eq:phi111}
	\varphi_\mathrm{e.d.} = 6 \lambda_{xyxy} \varepsilon_{xy}^2 .
\end{equation} 
The point symmetry is reduced to $32$, and six helicoids are distributed into two groups. Three helicoids perpendicular to the axis $[111]$ --- $[1\bar{1}0]$, $[01\bar{1}]$, $[\bar{1}01]$ --- do not change their rotation planes. The other three helicoids, with axes along the directions $[110]$, $[011]$, $[101]$, are connected by a three-fold rotation axis, and, consequently, have identical coefficients $\alpha$. Thus, the problem of minimizing elastic strain energy again reduces to determining a single parameter.

\begin{figure}[h]
	\begin{center}
		\includegraphics[width=8cm]{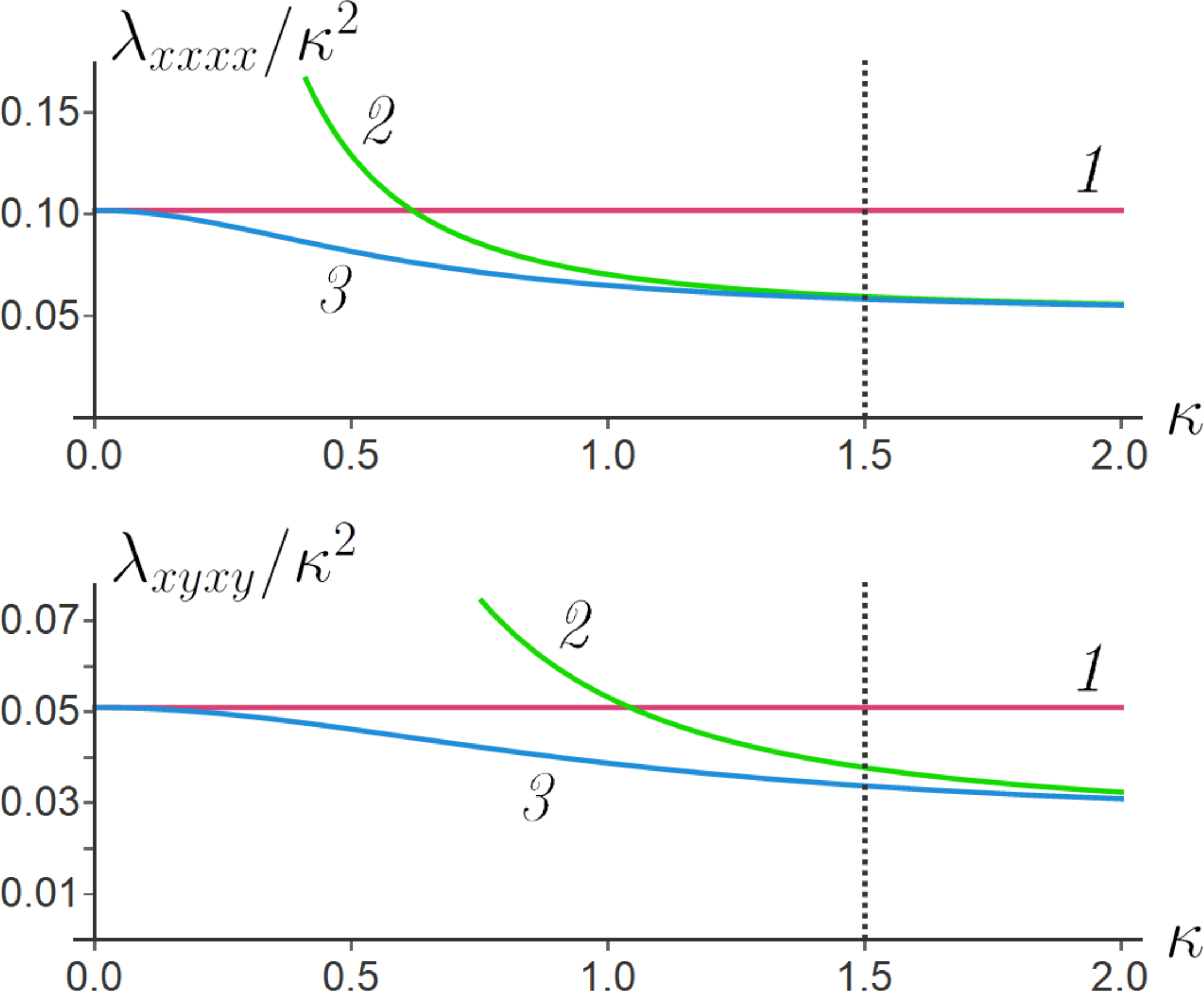}
		\caption{\label{fig:lambda} Dependence of components $\lambda_{xxxx}$ and $\lambda_{xyxy}$ of the blue phase $O^5$ elastic tensor on chirality $\kappa$ in one-parameter approximation ($\eta = 1$) at $\tau = \tau_c$: {\it 1} --- rigid tensor approximation, {\it 2} --- free helicoid approximation, {\it 3} --- interpolation model. For viewing convenience, all graphs are divided by $\kappa^2$. Parameter $\kappa = \tfrac32$ corresponds to the boundary between high and low chirality.}
	\end{center}
\end{figure}

\begin{figure}[h]
	\begin{center}
		\includegraphics[width=8cm]{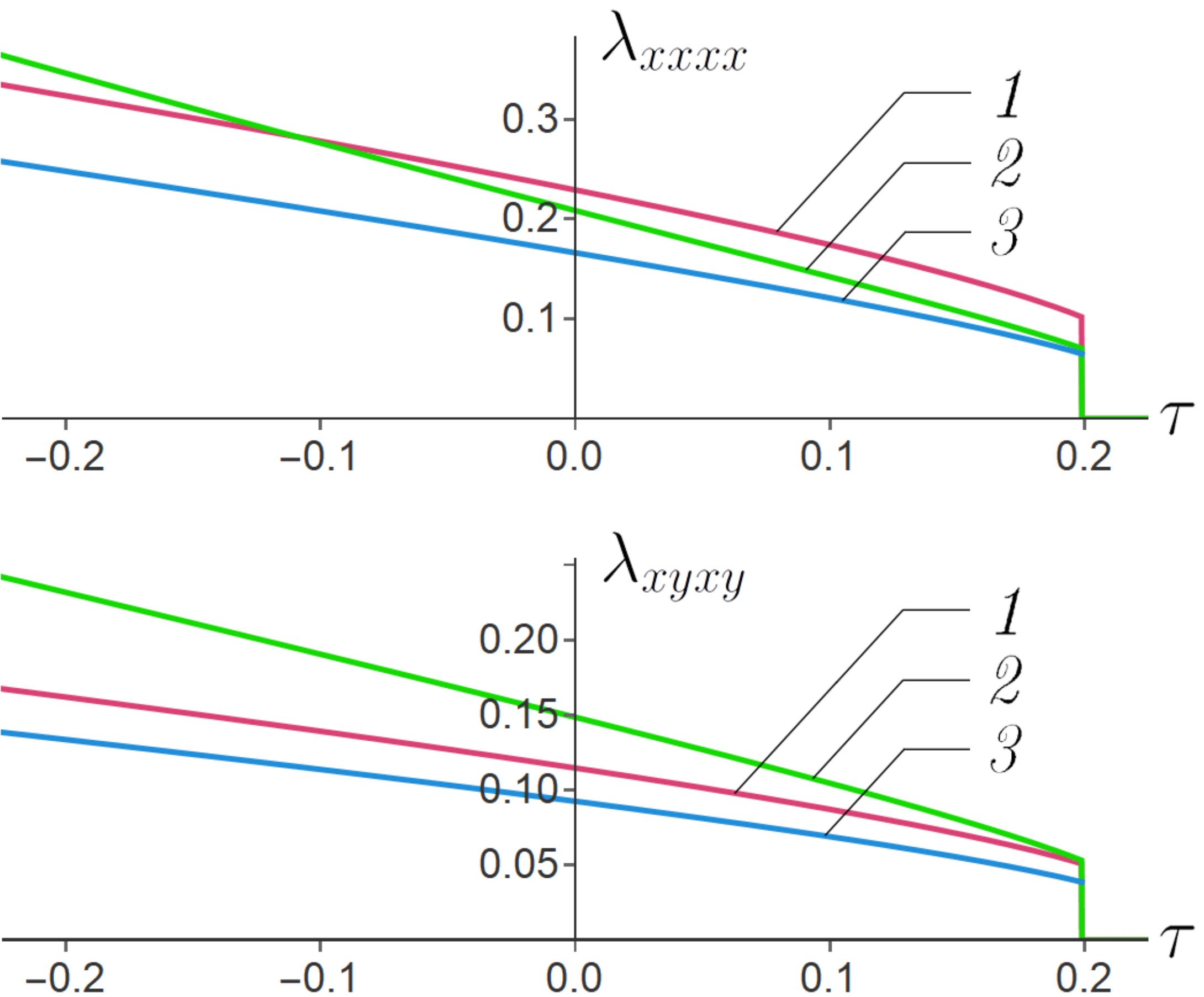}
		\caption{\label{fig:lambda2} Temperature dependence of components $\lambda_{xxxx}$ and $\lambda_{xyxy}$ of the blue phase $O^5$ elasticity tensor in one-parameter approximation ($\eta = 1$) at $\kappa = 1$: {\it 1} --- rigid tensor approximation, {\it 2} --- free helicoid approximation, {\it 3} --- interpolation model}
	\end{center}
\end{figure}

Routine calculations lead to the following expression for the elasticity tensor components $\hat{\lambda}$:
\begin{equation}
	\label{eq:synthetic}
	\lambda_\varpi = \lambda_{1,\varpi} - \frac{(\lambda_{1,\varpi} - \lambda_{2,\varpi})^2}{\lambda_{1,\varpi} - \lambda_{2,\varpi} + \Delta \lambda_\varpi} ,
\end{equation} 
where the index $\varpi$ takes values $xxxx$, $xxyy$, $xyxy$. For LC with low chirality ($\kappa \rightarrow 0$) this expression gives $\hat{\lambda} \approx \hat{\lambda}_1$, corresponding to the rigid tensor approximation. In the case of high chirality ($\kappa \rightarrow \infty$) we obtain the expression
\[
\hat{\lambda} \approx \hat{\lambda}_2 + \Delta \hat{\lambda}
\]
for free helicoids. Thus, this model connects the two previously considered approximations, and it is applicable for any values of the parameter $\kappa$. Fig.~\ref{fig:lambda} shows characteristic dependencies of the elasticity tensor components calculated for the one-parameter approximation ($\eta = 1$) near the transition point $\tau_c$. Example temperature dependence of elasticity tensor components is shown in Fig.~\ref{fig:lambda2}. It should be noted that equation (\ref{eq:synthetic}) does not have a tensor form, as it connects the corresponding tensor components in a nonlinear way. The latter fact may indicate some hidden symmetry not accounted for in our simple model.

\section{Conclusions}
\label{sec:conclusions}

The detailed examination of crystalline elastic properties of the simplest cubic blue phase $O^5$ in various models provided a comprehensive picture of changes in all three elastic constants at arbitrary temperature and chirality strength. Interesting special cases were identified: in the case of equal elastic constants in the microscopic Landau--de Gennes free energy ($\eta=1$) the blue phase in the rigid tensor approximation appears to be elastically isotropic, while in the free helicoid approximation, the Cauchy conditions are satisfied for the shear components of the elasticity tensor $\lambda_{xxyy}=\lambda_{xyxy}$. Future plans include applying the approaches developed in this article to calculate elastic constant tensors of actually observed blue phases $O^2$ and $O^8$ and hexagonal skyrmion structures in chiral magnetics. It would also be very interesting to account for anharmonic (nonlinear) contributions to elastic energy, which may prove significant due to the anomalously small elastic moduli in harmonic elasticity theory. Notable fluctuation phenomena characteristic of weak crystallization, which apparently lead to the experimentally observed blue fog phase BP~III, have not yet received quantitative theoretical description.

\section*{Acknowledgements}

The authors are grateful to M.~V.~Gorkunov, P.~V.~Dolganov, V.~K.~Dolganov, and E.~I.~Kats for critical remarks and useful discussions.

\section*{Funding}

This work was supported by the Russian Science Foundation, grant No. 23--12--00200. The numerical evaluation of elastic moduli was carried out by V.~A.~Chizhikov within the framework of the state assignment of the National Research Center ``Kurchatov Institute''.


\end{document}